\documentclass[fleqn,usenatbib]{mnras}

\usepackage{newtxtext,newtxmath}

\usepackage[T1]{fontenc}
\usepackage{ae,aecompl}

\usepackage{graphicx}	
\usepackage{amsmath}	
\usepackage{amssymb}	

\title[Wide binaries as probes of gravity]{The geometric challenge of testing gravity with wide binaries}

\author[K. El-Badry]{
Kareem El-Badry$^{1}$\thanks{E-mail: kelbadry@berkeley.edu}
\\
$^{1}$Department of Astronomy and Theoretical Astrophysics Center, University of California Berkeley, Berkeley, CA 94720
}

\date{Accepted to MNRAS}

\pubyear{2015}

\begin{document}
\label{firstpage}
\pagerange{\pageref{firstpage}--\pageref{lastpage}}
\maketitle

\begin{abstract}
Wide binaries provide promising laboratories for testing general relativity (GR) in the low-acceleration regime. Recent observational studies have found that the difference in the proper motions and/or radial velocities of the components of nearby wide binaries appear larger than predicted by Kepler's law's, indicating a potential breakdown of GR at low accelerations. These studies have not accounted for projection effects owing to the different position of the two stars on the celestial sphere. I show that two stars in a wide binary with {\it identical} 3D space velocities often have significantly different proper motions and radial velocities purely due to projection effects. I construct a sample of simulated binaries that follow Kepler's laws and have similar phase-space distributions to the observed samples of nearby binaries. Beyond separations of $\sim$\,0.1\,pc, direct comparison of the components' proper motions would suggest strong tensions with GR, even though the simulated binaries follow Kepler's laws by construction. The magnitude of the apparent disagreement is similar to that found observationally, suggesting that the apparent tension between observations and GR may largely be due to projection effects. I discuss prospects for constraining gravity at low accelerations with wide binaries. Robust tests of GR are possible with current data but require measurements of 3D velocities. Further work is also needed to model contamination from unbound moving groups and unrecognized hierarchical triples. 
\end{abstract}




\section{Introduction}

Wide binaries provide an intriguing testbed for modified theories of gravity that predict deviations from general relativity (GR) in the Newtonian limit at very low accelerations. Many such theories have been proposed to explain observations on galactic scales, potentially alleviating the need for dark matter to explain observations of galactic dynamics \citep[e.g.][]{Famaey_2012}. Testing modified gravity on galactic scales is challenging because many aspects of the galaxy formation process remain imperfectly understood. In the idealized case, the orbits of wide binaries provide a less complicated test, as the two-body problem has a simple solution both in GR and in many modified gravity theories \citep{Zhao_2010}. 

Modified gravity theories designed to explain galactic dynamics typically deviate from the Newtonian limit of GR at accelerations below the scale $a_0 \approx 10^{-8}\,\rm cm\,s^{-2}$. This is comparable to the acceleration in a solar-type binary with separation 10,000\,AU (0.05\,pc). At wider separations, GR and modified gravity theories predict different relations between the physical separation and orbital velocity of two stars in a gravitationally bound binary. The orbital timescales of such wide binaries are long ($\gtrsim$\,1 Myr), so their full 3D separation and orbital velocity are generally not directly observable. However, for a statistical sample of binaries observed at random snapshots in their orbits, GR and modified gravity theories predict different relations between the projected physical separation, $s$, and the instantaneous one-dimensional velocity difference between the two stars (typically measured with proper motions and/or radial velocities), $\Delta V$. The magnitude of the difference compared to GR varies significantly between modified gravity theories and is much larger in theories that do not include an external field effect than in theories that do \citep{Banik_2018, Pittordis_2018}. GR (in the Newonian limit) predicts $\Delta V \sim s^{-1/2}$; at $s\sim1$\,pc, it predicts bound solar-type binaries to have $\Delta V$ of only a few tens of $\rm m\,s^{-1}$. Measurements of binaries with wide separations and large $\Delta V$ can in principle rule out both GR and modified gravity theories that include an external field effect \citep{Famaey_2012, Pittordis_2018}.

\begin{figure*}
    \centering
    \includegraphics[width=\textwidth]{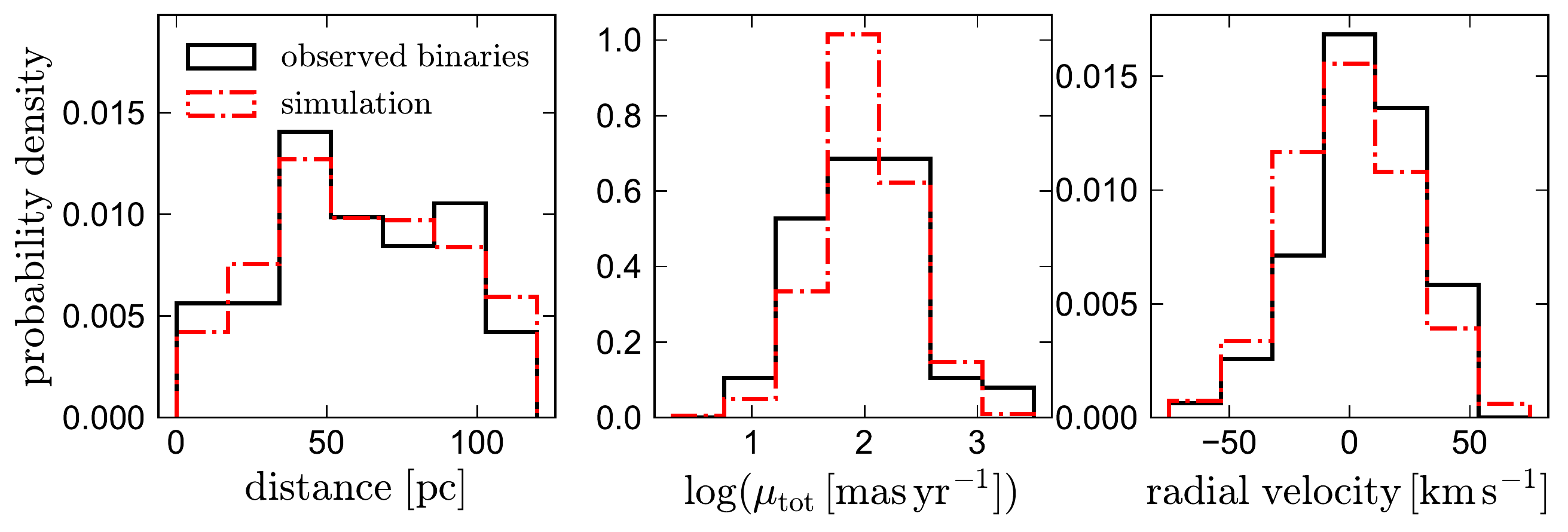}
    \caption{Distributions of heliocentric distance (left), total proper motion ($\mu_{{\rm tot}}=\sqrt{\mu_{\alpha}^{2}\cos^{2}\delta+\mu_{\delta}^{2}}$; middle) and radial velocity (right). The observed binary sample from \citet{Hernandez_2018} is shown in black; the simulated sample described in Section~\ref{sec:sims} is shown in red. The simulated sample is constructed to have a similar distribution of distance and proper motion to the observed sample.}
    \label{fig:demographics}
\end{figure*}

Recently, \citet{Hernandez_2018} used astrometry from {\it Gaia} for a sample of wide binary candidates with projected separations $0.01 \lesssim s/{\rm pc} \lesssim 10$ to measure the relation between $\Delta V$ and $s$, with the aim of testing classical gravity at low accelerations. They found values of $\Delta V$ that are substantially larger than the Newtonian prediction at large separations ($s\gtrsim 0.1$\,pc) and interpreted them as evidence for a possible breakdown of GR at low accelerations. Their binary catalog contains 83 wide binary candidates within $\sim$\,120 pc of the Sun that were originally identified by \citet{Shaya_2011} using astrometry from the {\it Hipparcos} and {\it Tycho-2} catalogs and were classified as unlikely to be chance alignments of stars that are not physically associated. The second {\it Gaia} data release significantly improved the precision of proper motions and parallaxes for most stars observed by {\it Hipparcos}, making it possible to measure sky-projected velocities for nearby stars with $\ll 0.1\,\rm km\,s^{-1}$ precision \citep{Gaia_2018}. As a result, the formal constraints on $\Delta V$ obtained by \citet{Hernandez_2018} are quite strong.

Prior to the {\it Gaia} mission, two other studies considered the relation between $\Delta V$ and $s$ as a potential constraint on the force law at low accelerations. \citet{Hernandez_2012} used proper motions of wide binary candidates from the SDSS ``SLoWPoKES'' catalog \citep{Dhital_2010}, as well as those of some of the wide binary candidates in the \citet{Shaya_2011} catalog, and reached qualitatively similar conclusions to \citet{Hernandez_2018}, but with larger uncertainties due to less precise astrometry. \citet{Scarpa_2017} compared the radial velocities (RVs) of the components of some of the nearby binaries studied by \citet{Hernandez_2012}. They found that a large fraction had RV differences larger than the naive prediction for Keplerian orbits, indicating either a breakdown in classical gravity or that the binary candidates are not actually gravitationally bound.

In this paper, I draw attention to a potential complication in measuring the true velocity differences of the components of wide binaries. Direct comparison of RVs and proper motions of binary components entails the projection of velocity vectors in spherical coordinates onto a local Cartesian frame centered on each component. This will yield correct velocity differences in the limit where the two components are at the same position on the sky. However, as the angular separation of binary components grows, Cartesian planes that are normal to the unit sphere at the position of each component become rotated with respect to each other, leading to projection effects \citep[e.g.][]{Shaya_2011}. These effects can cause two stars with very similar or identical 3D space velocities to have significantly different proper motions and radial velocities. Left unaccounted for, projection effects can lead to apparent tension with GR at wide separations, even in the absence of any true deviation from Keplerian orbits. In Section~\ref{sec:sims} below, I describe simulations to quantify the magnitude of this effect. 

\section{Simulations}
\label{sec:sims}

I construct a sample of simulated binaries with the goal of comparing to the sample studied by \citet{Hernandez_2018}. I first sample positions for the center of mass of each binary assuming a uniform spatial distribution. I then reject a random subset of the simulated binaries such that their distribution of heliocentric distance is similar to that of the observed sample. Center-of-mass velocities for each simulated binary are drawn from a 3D Gaussian with $\sigma_{\rm 1D}=25\,{\rm km\,s^{-1}}$. This value of $\sigma_{\rm 1D}$ is comparable to that  measured in the solar neighborhood \citep{Sharma_2014} and is also equal to the dispersion in RV for the observed sample of binaries that have measured RVs. In Figure~\ref{fig:demographics}, I compare the distributions of distance, proper motion, and RV of the simulated binary sample to those of the observed sample. 

For each simulated binary, I draw an eccentricity from a uniform distribution $e\in [0, 1]$, a primary mass from a uniform distribution of $m_1 \in [0.5M_{\odot}, 0.6M_{\odot}]$, and a mass ratio from a uniform distribution of $q = m_2/m_1 \in [0.8, 1]$. These choices are designed to yield a binary population similar to the observed sample. I assume random orbital orientations, corresponding to a $p(i)\,{\rm d}i = \sin(i)\,{\rm d} i$ distribution of inclinations. Periods are drawn from a log-uniform distribution. This is not realistic \citep{Duchene_2013}, but the period distribution is not important for this study, which aims only to measure the enhancement of $\Delta V$ {\it at a particular separation}. Finally, each binary is mock observed at a random time $t_{\rm obs} \in [0, P]$. 

At time $t_{\rm obs}$, I calculate the celestial coordinates of each component star, and $\theta$, the angular separation between them. For a binary with center-of-mass distance $d$, the projected separation is $s=d\times \theta$, and the projected physical velocity difference in each proper motion component is 
\begin{align}
\label{eq:delta_v}
    \frac{\Delta V_{i}}{{\rm km\,s^{-1}}}=4.74\times10^{-3}\frac{\Delta\mu_{i}}{{\rm mas\,yr^{-1}}}\times\frac{d}{{\rm pc}}. 
\end{align}
Here $\Delta\mu_{i}$ represents the difference in each proper motion component ($\Delta \mu_{\alpha}\cos(\delta)$ for right ascension; $\Delta \mu_{\delta}$ for declination) between the two components of a binary. I also calculate the RV of each component, and the true velocity difference between the components of each binary along the three Cartesian axes. Transformations between coordinate frames are carried out using the \texttt{astropy.coordinates} package \citep{Astropy_2018}.

\subsection{Projection effects}
\label{sec:proj_effects}

\begin{figure}
    \centering
    \includegraphics[width=\columnwidth]{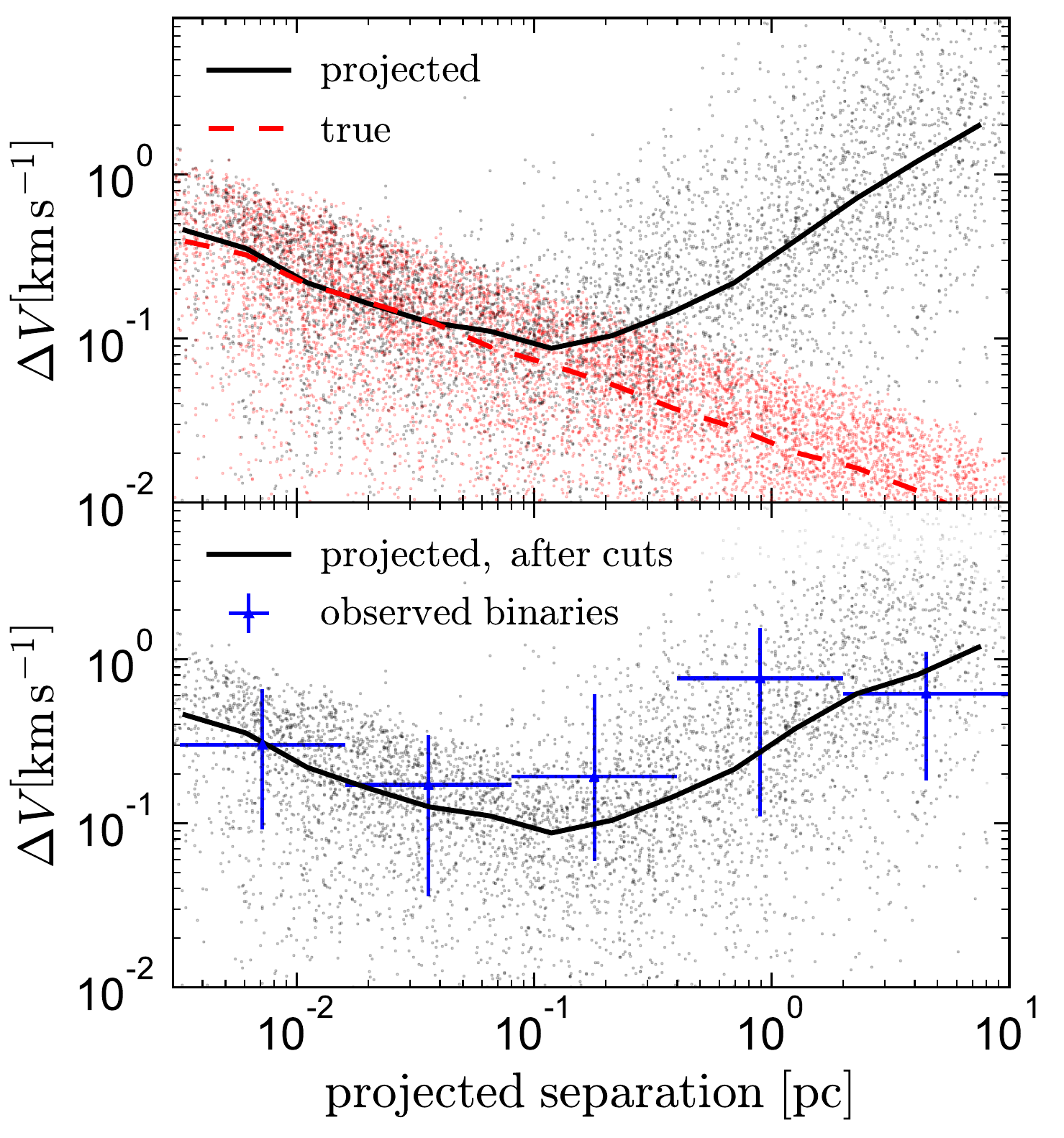}
    \caption{One-dimensional velocity difference versus projected separation for simulated binaries. In the top panel, red points show the true velocity difference, while black points show the result of directly comparing proper motion components; i.e. applying Equation~\ref{eq:delta_v}. Solid and dashed lines show the median $\Delta V$ in bins of projected separation. At $s \gtrsim 0.1\,\rm pc$, projection effects become important, and the value of $\Delta V$ obtained by direct comparison of proper motion components significantly exceeds the true velocity difference. In the bottom panel, I remove binaries in which $\Delta V >4\rm \,km\,s^{-1}$ in any component, as such pairs are unlikely be classified as binaries in the first place. Blue points and error bars show the binned median and middle 68\% of $\Delta V$ for the observed binary sample from \citet{Hernandez_2012}. The increase in $\Delta V$ at large separations is quite similar to that predicted to result from projection effects in the simulations. }
    \label{fig:proj_effects}
\end{figure}

Figure~\ref{fig:proj_effects} compares the true one-dimensional $\Delta V$ to the value obtained by directly comparing the proper motions of the two componets (Equation~\ref{eq:delta_v}). Each binary contributes two points, as proper motions in the RA and Dec components are considered independently. At small separations, projection effects are negligible, and the median $\Delta V$ computed from Equation~\ref{eq:delta_v} is identical to that of the true 1D velocity difference computed from the Cartesian velocity components. However, at $s\gtrsim 0.1$\,pc (corresponding to an angular separation of $\theta\gtrsim 10$ arcminutes for the distance distribution of our sample), the true and projected values of $\Delta V$ begin to diverge widely. At a separation of 1 pc (10 pc), the typical enhancement in $\Delta V$ due to projection effects is 0.5 km\,s$^{-1}$ (2 km\,s$^{-1}$).

The bottom panel of Figure~\ref{fig:proj_effects} compares the values of $\Delta V$ predicted by the simulation to the observed binary sample from \citet{Hernandez_2018}. Here I have removed binaries in which the RVs or either proper motion component of the two stars differ by more than 4 km\,s$^{-1}$; such pairs are unlikely to be classified as genuine binaries.\footnote{This is similar but not identical to the cuts used in constructing the observed binary sample studied by \citet{Hernandez_2018}. Their sample was constructed from the \citet{Shaya_2011} binary catalog, after removal of pairs in which both components have RVs measured by {\it Gaia} that differ by more than 4\,km\,s$^{-1}$, as well as pairs in which \citet{Shaya_2011} estimated the probability that the system is a chance alignment to be greater than 10\%. 

\citet{Shaya_2011} did attempt to account for geometrically-induced proper motion differences in constructing their catalog. Because \citet{Hernandez_2012} used measurements of $\Delta \mu$ from \citet{Shaya_2011} for part of their sample, these measurements may be less affected by projection effects than those in \citet{Hernandez_2018}.} I plot the median and 1$\sigma$ (middle 68\%) range of $\Delta V$ for observed binaries in each bin of projected separation (note that this is not identical to the rms velocity in each bin plotted by \citealt{Hernandez_2018}). Values of $\Delta V$ for the observed binaries are generally in good agreement with the simulation, even at large separations where they disagree substantially from the prediction for bound Keplerian orbits. This agreement suggests that the tension between the observed sample and the classical gravity prediction at large separations may in large part be a consequence of projection effects.  

Although it is not shown in Figure~\ref{fig:proj_effects}, I find that projection effects enhance $\Delta V$ in the RV component in a manner essentially identical to the enhancement in proper motion difference. 

\subsection{Correcting for projection effects}
\label{sec:correction}

\begin{figure}
    \centering
    \includegraphics[width = \columnwidth]{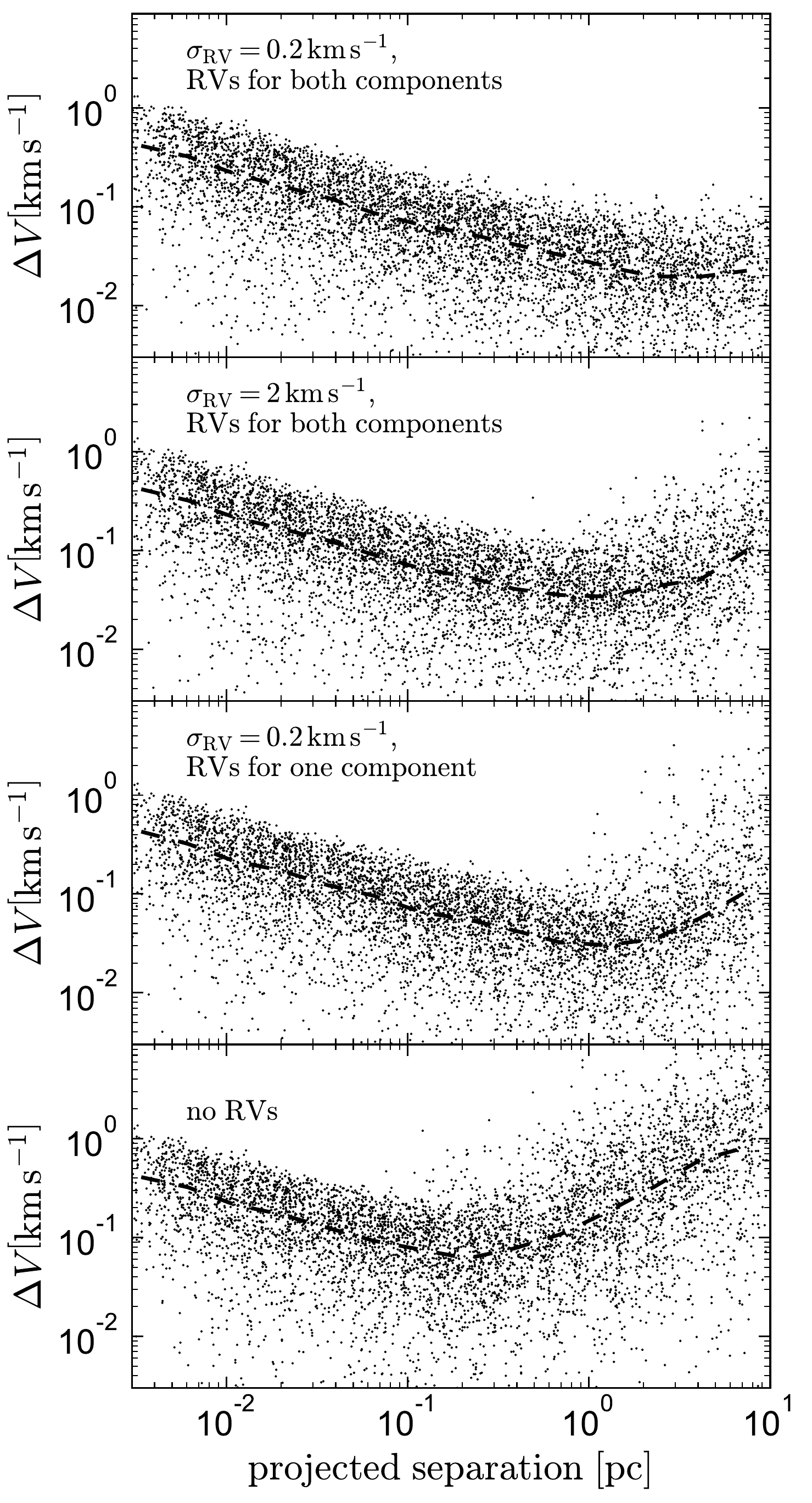}
    \caption{Predicted one-dimensional $\Delta V$ recovered from precise proper motions when projection effects are corrected for (see Section~\ref{sec:correction}). In the top two panels, I assume RVs are available for both components. The third panel assumes RVs available for only one component, and the fourth panel assumes that no RVs are available for either component. Correcting for projection effects ameliorates the enhancement in $\Delta V$ at large separations compared to the direct comparison of proper motions in all cases, but some RV information is required to recover accurate velocities in the separation regime where alternative gravity theories predict strong deviations from GR.}
    \label{fig:3d_vels}
\end{figure}

With measurements of both radial velocity and proper motion, projection effects can be straightforwardly corrected for. The most direct approach is to transform the velocities of both components into a Cartesian frame and compare each of the velocity components in that frame. With {\it Gaia} data for nearby binaries, it is typically the case that proper motion uncertainies are much smaller than RV uncertainties. This fact can be exploited by transforming velocities for both stars into a Cartesian frame centered on one of the components and normal to the celestial sphere (i.e., with the $\hat{z}$ axis pointing toward the Sun). In this case, a high-precision comparison of velocities in the $\hat{x}$ and $\hat{y}$ components is possible even with relatively low-precision RVs. 

This is demonstrated in Figure~\ref{fig:3d_vels}, which shows the $\Delta V$ predicted for simulated binaries after transforming to a Cartesian frame centered on one component. I assume a proper motion uncertainty of $5\,\rm mas\,yr^{-1}$ in both components in all cases (adding Gaussian noise to the proper motions during mock observations) and vary the available RV information in different panels. To correct for projection effects, I chose one component of each binary at random and transform the 3D velocities of both stars to a Cartesian frame that is centered on that component with $\hat{z}$ axis pointed toward the Sun. I calculate $\Delta V$ separately for the $\hat{x}$ and $\hat{y}$ components in this frame. If no RVs are available, I assume RV = 0\,km\,s$^{-1}$ for both components. If an RV is measured for one component only, I assume both component have that same RV.

When no RVs are available (bottom panel), transforming to a Cartesian frame reduces the enhancement of $\Delta V$ due to projection effects by a factor of $\sim$\,2 on average, but projection effects still dominate at $s \gtrsim 0.2$\,pc. Inclusion of RVs for both components makes the geometric correction effective to typical separations of 1-3 pc, depending on the precision of the RVs (top two panels). Obtaining a precise RV for one component and assuming that the 2nd component has the same RV is also effective at $s \lesssim 1$\,pc.

\section{Discussion}

I have shown that projection effects can lead to apparent disagreement in the proper motions and RVs of the components of wide binaries, even when the 3D space velocities follow Keplerian orbits. The magnitude of the disagreement caused by projection effects is comparable to the disagreement found observationally \citep{Scarpa_2017, Hernandez_2018}. It is of course possible that the observed wide binary candidates really do have discrepant 3D velocities, even after geometric distortions are corrected for. Here I only argue that consistency with Keplerian orbits is not ruled out. 

I have considered only geometric complications to attempts to measure the velocity difference between the two components of a binary. Below, I briefly highlight two additional challenges to tests of gravity with wide binaries, and their potential resolutions.  

\begin{enumerate}
    \item {\it Unbound comoving pairs}: There is no guarantee that wide pairs of comoving stars with consistent astrometry are gravitationally bound. Recent studies \citep[e.g.][]{Oh_2017, Simpson_2018, Faherty_2018} have identified large numbers of unbound ``moving groups''; i.e., associations of stars that are likely not  gravitationally bound but follow very similar orbits through the Galaxy. Moving groups originate from dissolving star clusters that drift apart on $\sim$\,100 Myr timescales. Their components have typical separations of a few pc and are essentially indistinguishable from genuinely bound binaries at large separations. 
    
    Failure to remove unbound pairs will result in an artificial enhancement of $\Delta V$ at large separations. Unbound pairs are more common at large separations, where the number of genuine binaries decreases, the phase-space volume for potential companions increases, and the binding energy of genuine binaries decreases. At separations exceeding the Jacobi limit, $r_{\rm J} \sim1.7$\,pc, the Galactic tidal field is stronger than the gravitational attraction between the components of solar-type binaries \citep{Binney_2008}, so essentially no binaries at larger separations are expected to be genuinely bound. Several previous studies have attempted to account for the presence of unbound pairs by comparing to the predictions of \citet{Jiang_2010}, who include in their calculations disrupted, unbound binaries in the process of drifting gradually apart. But because \citet{Jiang_2010} model only co-moving pairs that were initially gravitationally bound binaries -- i.e., their calculations do not attempt to account for the fact that most stars are born in clusters that gradually dissolve -- their prediction at large separations likely underestimates the number of unbound comoving pairs with velocities that differ by a few km\,s$^{-1}$.
    
    At the widest separations where previous studies have attempted to use wide binaries as probes of gravity, it is highly improbably that the identified pairs are binaries in any useful dynamical sense. At a separation of 8 pc, typical orbital periods are $\sim$1.5 Gyr, many times larger than the Galactic dynamical time! It is therefore advisable to use binaries with separations $s < 1$\,pc. The rate of contamination from dissolving clusters can also be substantially decreased by targeting binaries on halo-like orbits, which are primarily old and free from their birth associations. 
    
    \item {\it Hierarchical triples}: nearly half all wide binaries are really hierarchical triples and higher-order multiples \citep{Tokovinin_2006, ElBadry_2018} in which the additionally components are either faint or unresolved. These systems are generically not expected to follow the Keplerian prediction of $\Delta V \sim s^{-1/2}$, because the dynamics of one component are dominated by its closer companion. Of order half of such systems can be identified as containing photometric or spectroscopic binaries \citep{ElBadry_2018b}, but it is unavoidable that a substantial fraction of hierarchical triples will go undetected; namely, those in which the unseen companion is faint and the inner orbit is wide enough that little RV variation is expected on observable timescales. Robust tests of gravity with samples of wide binaries should therefore include a population model for unrecognized higher-order multiples. 
    
\end{enumerate}

\section*{Acknowledgements}

I thank Xavier Hernandez, Andrian Price-Whelan, Eliot Quataert, and Daniel Weisz for useful discussions and am grateful to the anonymous referee for a constructive report.
I acknowledge support from an NSF graduate research fellowship. 



\bibliographystyle{mnras}




\bsp	
\label{lastpage}
\end{document}